\documentclass[fleqn,usenatbib]{mnras}

\usepackage{newtxtext,newtxmath}

\usepackage[T1]{fontenc}

\DeclareRobustCommand{\VAN}[3]{#2}
\let\VANthebibliography\thebibliography
\def\thebibliography{\DeclareRobustCommand{\VAN}[3]{##3}\VANthebibliography}

\usepackage{graphicx}	
\usepackage{amsmath}	

\title[Microlensing in quasar broad emission lines]{Evidence for
 microlensing by primordial black holes in quasar broad emission lines}

\author[M. R. S. Hawkins]{
M. R. S. Hawkins $^{1}$\thanks{E-mail: mrsh@roe.ac.uk}
\\
$^{1}$Institute for Astronomy (IfA), University of Edinburgh,
 Royal Observatory, Blackford Hill, Edinburgh EH9 3HJ, UK\\}

\date{Accepted XXX. Received YYY; in original form ZZZ}

\pubyear{2022}

\begin{document}
\label{firstpage}
\pagerange{\pageref{firstpage}--\pageref{lastpage}}
\maketitle

\begin{abstract}
With the detection of black hole mergers by the LIGO gravitational wave
telescope, there has been increasing interest in the possibility that dark
matter may be in the form of solar mass primordial black holes.  One of
the predictions implicit in this idea is that compact clouds in the broad
emission line regions of high redshift quasars will be microlensed,
leading to changes in line structure and the appearance of new emission
features.  In this paper the effect of microlensing on the broad emission
line region is reviewed by reference to gravitationally lensed quasar
systems where microlensing of the emission lines can be unambiguously
identified.  It is then shown that although changes in Seyfert galaxy
line profiles occur on timescales of a few years, they are too nearby for
a significant chance that they could be microlensed, and are plausibly
attributed to intrinsic changes in line structure.  In contrast,
in a sample of 53 high redshift quasars, 9 quasars show large changes in
line profile at a rate consistent with microlensing.  These changes occur
on a timescale an order of magnitude too short for changes associated with
the dynamics of the emission line region.  The main conclusion of the
paper is that the observed changes in quasar emission line profiles are
consistent with microlensing by a population of solar mass compact bodies
making up the dark matter, although other explanations like intrinsic
variability are possible.  Such bodies are most plausibly identified as
primordial black holes.
\end{abstract}




\section{Introduction}
\label{int}

The recent detection of black hole mergers by the LIGO gravitational wave
observatory has in the first instance been attributed to the merging of
massive stellar black hole remnants \citep{a16}.  However, it has
also been seen as adding considerable weight to the idea that dark matter
is in the form of primordial black holes \citep{b16}.  The idea was
that the detections would form part of a high mass tail to an expected
broad mass function peaking at around a solar mass.  From a theoretical
perspective, a mechanism for the formation of primordial black holes in
the early Universe was discussed by \cite{c74}, with the additional
suggestion \citep{c75} that such objects might make up the dark matter.
Constraints appearing to rule out this idea on the basis that it implied
excessive, unseen variations in quasar brightness due to microlensing by
the black holes acting as lenses \citep{s93} were shown to be insecure by
\cite{z03}, on the basis that an unrealistically small source size had
been assumed for the quasar accretion disc.

The first claim that primordial black holes had actually been detected
\citep{h93} made the case that the observed brightness variations in
samples of quasars could only be explained by the effects of microlensing
by a large population of compact bodies, most plausibly primordial black
holes.  These bodies would have to make up a large fraction of the dark
matter.  In the ensuing years, a number of varied observations suggested
that optical variability in quasars was at least partly the result of
microlensing by a population of stellar mass compact bodies.  These
results were summarised by \cite{h11} as a case for primordial black holes
as dark matter.

The idea that dark matter is in the form of compact bodies has a long
and controversial history.  An early review by \cite{t87} largely focussed
on a variety of elementary particles as dark matter candidates, although
she did include a number of compact objects including cosmic strings,
quark nuggets and primordial black holes as alternative possibilities.
It seems fair to say that the consensus view at that time was that dark
matter was in the form of elementary particles, which would soon be
detected by one of a number of ongoing experiments.  In the event,
particle dark matter has not so far been detected, and as detection limits
approach the neutrino floor, the prospects for such detections are not
good.  There have nonetheless been a number of attempts to detect or put
limits on any population of compact bodies which might contribute to or
account for the dark matter.  Perhaps the most significant of these was
the large scale survey by the MACHO collaboration \citep{a00} to detect
microlensing events in the light of Magellanic Cloud stars by compact
bodies along the line of sight in the Galactic halo.  The results of this
project were controversial, as although the number of events detected
exceeded any known stellar population, it was less than that expected
for a halo entirely composed of compact bodies.

Since the publication of the results of the MACHO collaboration
\citep{a00}, there have been a number of attempts to constrain any
population of compact bodies making up the dark matter.  In the first
instance the microlensing observations were repeated by two other groups,
the EROS and OGLE collaborations \citep{t07,w11}.  The results of these
two surveys were in significant ways inconsistent with those of the MACHO
collaboration \citep{h15}, which led to a variety of new approaches to
constraining dark matter in the form of compact bodies, and in particular
primordial black holes \citep{c10,c17}.  The constraints included
so-called pixel-lensing in M31, brightness changes in Type Ia supernovae
due to microlensing, the disruption of wide binary star systems by compact
bodies, the depletion of stars in the centre of dwarf galaxies due to mass
segretation, excess X-ray luminosity in the Galactic centre due to
interaction of compact bodies with the inter-stellar medium and
distortions in the Cosmic Microwave Background due to accretion onto
primordial black holes in the early Universe.  However, all these
constraints are based on assumptions which have been vigorously challenged
in the literature \citep{b18}.

A more direct approach to the question of whether the dark matter in
galaxy halos can be in the form of compact bodies is based on analysing
photometric variations in the multiple images of gravitationally lensed
quasars \citep{m09,p12}.  The idea is that although intrinsic variations
in the quasar will be observed in all the quasar images, subject to time
delays approropriate to the light travel time to the individual images, it
is also the case that the images vary independently of each other.  This
is widely interpreted to be due to microlensing by a population of stellar
mass compact bodies, where the light from each quasar image traverses a
different amplification pattern on its trajectory to the observer.  The
mass estimate is derived from the characteristic timescale of the events,
and the question of interest is whether the lenses are stars in the
lensing galaxy, or compact bodies making up the dark matter halo.

The conventional approach to determining the stellar fraction of the
lensing galaxy halo has been to use a maximum likelihood estimate of the
ratio of mass in compact bodies to that in smoothly distributed particles,
based on the observed microlensing amplifications.  This procedure has
tended to give low values for the ratio of compact to smoothly distributed
matter, consistent with the stellar population of the lensing galaxy
acting as the lenses \citep{m09,p12}.  However, this result does not agree
with direct measurements of the stellar population in the vicinity of the
quasar images from the distribution of starlight in a sample of wide
separation systems where the quasar images lie well clear of the stellar
population, and yet are strongly microlensed \citep{h20a,h20b}.  The
problem seems to be that the maximum likelihood estimates are based on
large samples of mostly compact lens systems, where the quasar images are
buried deep within the stellar distribution of the lensing galaxy and the
disc stars form a large optical depth to microlensing.  In this case there
is no reason to doubt that the observed microlensing can be produced by
the stellar population.  This is clear from the work of \cite{p12}, where
in a subsample of wide separation lensed quasars the observed variations
are not consistent with microlensing by the relatively sparse stellar
population.

A direct measurement of star light in wide separation lensed quasar
systems where the images lie well clear of the stellar population
of the lensing galaxies \citep{h20a} has shown that stars in the galaxy
halos are far too sparse to account for the observed microlensing
amplifications.  The most convincing evidence comes from the cluster lens
SDSS J1004+4112 \citep{h20b} where strong microlensing is observed in the
light curves of the quasar images some 60 kpc from the cluster centre.  As
the optical depth to microlensing $\tau_*$ by the stellar population has
already dropped to negligible levels 25 kpc from the cluster centre, it
seems clear that the microlenses must be part of some other population of
compact bodies.  In addition, for quasars in the general field it has
been shown that the observed variations in quasar brightness cannot be
accounted for by intrinsic changes in luminosity \citep{h22}. The
additional contribution of the microlensing amplification predicted for a
population of solar mass compact bodies making up the dark matter is
required to provide a good match to the data.  There are a number of
constraints on the identification of a population of compact bodies making
up the dark matter \citep{h20a}.  For a start they must be non-baryonic,
as well as sufficiently compact to act as lenses, and the timescale of
microlensing events implies that the mass of the lenses must peak at
around a solar mass.  These constraints appear to rule out all known
candidates for the compact bodies apart from primordial black holes
\citep{h20a}.  In addition, there is a strong theoretical framework for
the creation of primordial black holes in the early Universe \citep{b18},
with a mass function peaking at around a solar mass.

For a quasar accretion disc to be microlensed, it must be more compact
than the Einstein ring associated with the lensing objects.  For most
cosmological situations this implies a lens of around a solar mass for a
typical quasar accretion disc.  The question of whether the broad emission
lines are microlensed is less clear, and depends on the structure of the
broad emission line region (BLR).  At the time of early work on
microlensing \citep{s90} it was generally believed that the BLR was large,
of the order of a light year, and an order of magnitude larger than a
typical Einstein ring for a solar mass lens.  On this basis, \cite{s90}
concluded that the flux from the BLR would not be significantly affected
by microlensing unless the internal structure of the BLR was non-uniform.
The idea that the BLR might not be uniform, but confined by magnetic
stresses \citep{r87} was developed by \cite{b01} to argue that the clouds
within the BLR are not isolated individual entities embedded within a
confining medium, but transient knots of higher density within an overall
turbulent BLR.  This idea was developed by \cite{l06} who showed that
although the integrated microlensing effect of a fractal structure would
result in an overall constant light curve, significant magnification of
substructures could alter the emission line profiles in the BLR.

\begin{figure*}
\centering
\begin{picture} (0,0) (255,10)
\includegraphics[width=0.49\textwidth]{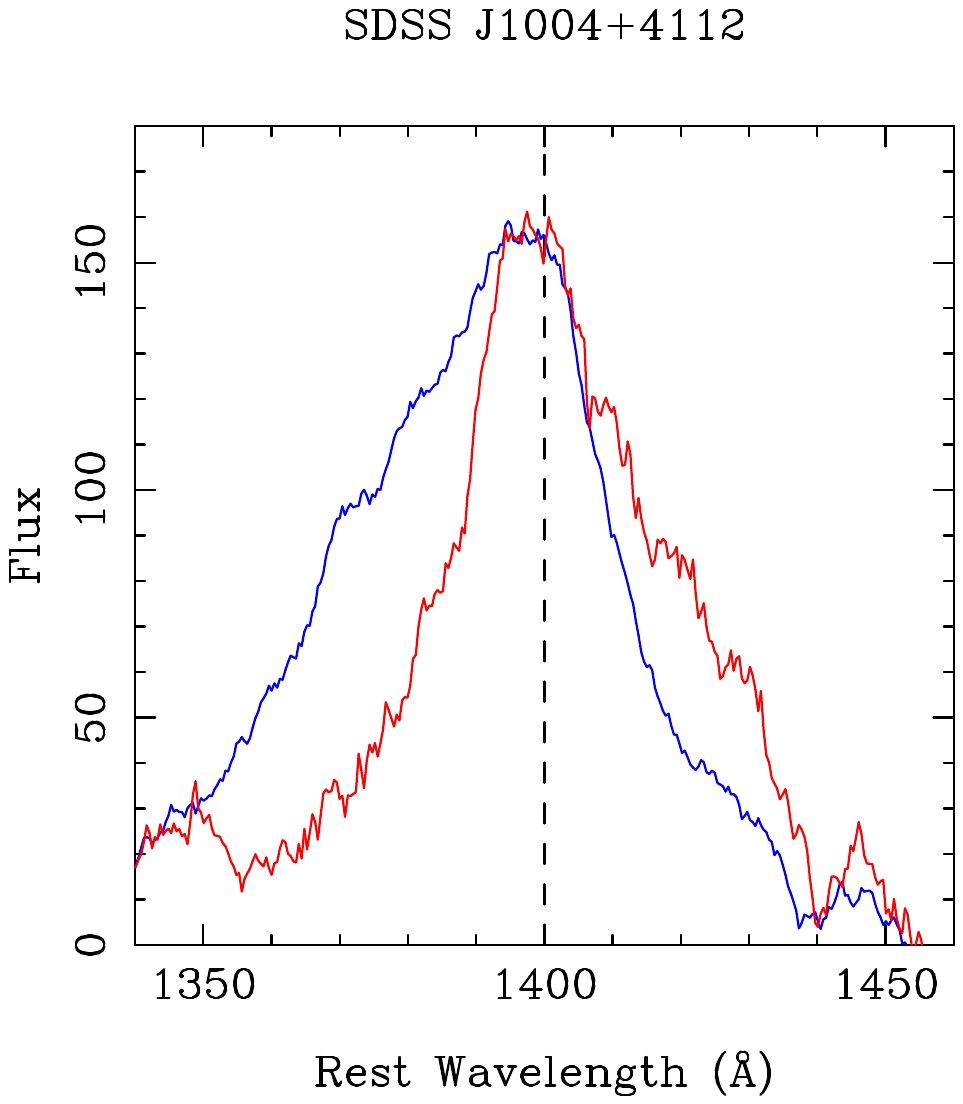}
\end{picture}
\begin{picture} (0,220) (-5,10)
\includegraphics[width=0.49\textwidth]{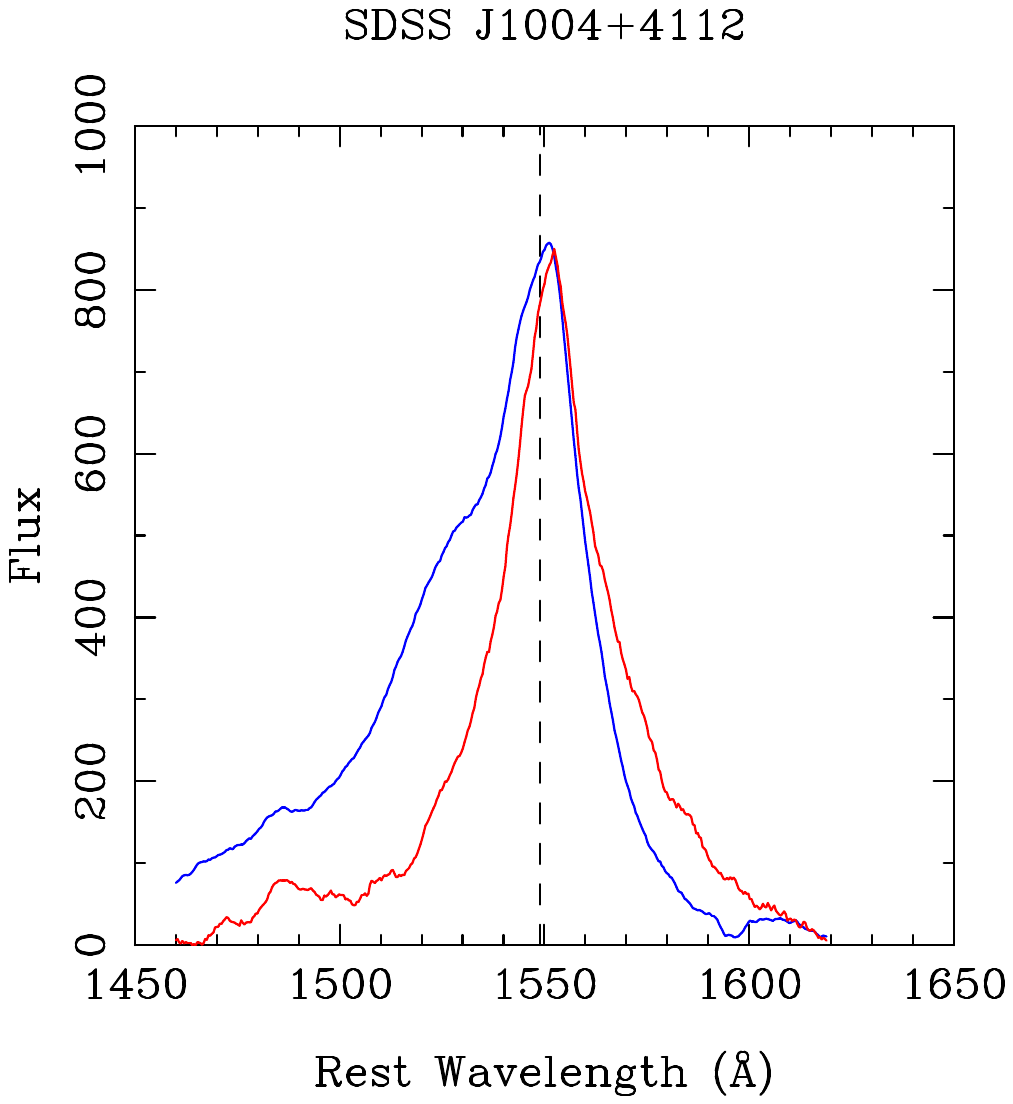}
\end{picture}
\caption{Emission lines from Keck spectra of the wide separation
gravitational lense system SDSS J1004+4112.  The left hand panel shows
the Si \textsc{iv} $\lambda1400$ line for the A image (blue), and the
B image (red).  The dashed vertical line shows the systemic wavelength.
The right hand panel shows similar data for the
C \textsc{iv} $\lambda1549$ line.}
\label{fig1}
\end{figure*}

One of the first gravitational lens systems to be analysed for
microlensing effects is the quadruply imaged Q2237+0305, also known as the
Einstein Cross.  A relative change in the brightness of two of the images
was definitively observed by \cite{i89}, and was in fact the first
detection of a microlensing event.  Q2237+0305 has continued to be a
valuable laboratory for exploring the effects of microlensing of the
continuum source associated with the accretion disc, and was an obvious
choice for early investigations into the possible microlensing of the BLR.
The first such study \citep{w05} was largely focussed on measuring the
ratio of the size of the C \textsc{iii}]/Mg \textsc{ii} regions.  In the
process they convincingly showed that the broad lines were being
microlensed.  They noted that the flux ratios for the two BLRs were
consistent with each other, but not with that for the continuum.  From
this they concluded that the two emission regions were of the same size,
and located along the same line of sight.

Since this early work there have been extensive efforts to look for
the effects of microlensing on broad line regions in gravitationally 
lensed quasar systems \citep{s12,g13b,f21}.  From these studies it is clear
that in most of these systems the broad line region is being microlensed.
This conclusion is primarily based on the observation that after allowing
for time delays beween the images, the structure of the emission lines is
significantly different.  This is interpreted as a consequence of the
differing amplification patterns due to a population of stellar mass
lenses traversed by light rays from each image to the observer.  The
question of the nature of the lenses is not addressed in these papers,
but the implication seems to be that they must be stars in the lensing
galaxy, as advocated in earlier work on microlensing of the continuum
light from the quasar accretion disc \citep{m09,p12}.  This orthodoxy has
recently been challenged \citep{h20a,h20b}, where it is shown that in wide
separation lens systems the stellar population of the lensing galaxy in
the vicinity of the quasar images is far to sparse to be responsible for
the observed microlensing.

The idea behind the present paper is as follows.  If the dark matter is
indeed made up of stellar mass primordial black holes, then these compact
bodies should microlens the broad emission lines of a substantial
fraction of quasar spectra.  This microlensing effect would result in a
change in the structure of the emission lines over a period of a few
years, by analogy with changes observed in the images of gravitationally
lensed quasars.  If no such changes are seen then this would be
inconsistent with the view that the dark matter is largely composed of
compact bodies.  If changes in emission line structure are observed, then
although these observations would be consistent with a compact body
component of the dark matter, there remains the question of whether they
can be attributed to intrinsic changes in the velocity structure of the
broad line region.

\section{Microlensing of the quasar broad line region}
\label{mic}

The question of the extent to which the quasar BLR can be microlensed is
an interesting one.  Firstly, the mass of the lensing bodies is of
particular importance as the maximum source size which can be
significantly microlensed is a function of the Einstein radius of the
lens.  For this investigation a lens mass of $1 M_\odot$ is adopted,
in line with the detections of solar mass bodies in the Galactic halo by
the MACHO collaboration \citep{a00}.  This implies an Einstein radius $R_E$
of the order of $10^{17}$cm, or 40 lt-day for a typical lensed quasar
system.  Early measures of the size of the BLR in quasars suggested an
upper limit of around 1 lt-yr, but it soon became clear from a large
sample of quasars monitored for several years \citep{k00} that typical
rest frame time lags were around 100 days.  This is close enough to the
size of $R_E$ to expect measurable microlensing effects as the lenses
cross the line of sight to the BLR.

In order to detect microlensing in the BLR, a strategy similar to that
used for microlensing of the accretion disc has been employed.  The idea
here is to monitor the spectra of individual images in gravitationally
lensed quasar systems, and after allowing for differences in light travel
time to compare the strength and shape of the broad emission lines.  As
all images are of the same quasar, corrected to the same epoch, any
differences are taken to be the result of microlensing \citep{r04,w05}.
In the event that the BLR is smaller than the Einstein radius of the lens,
the effect of microlensing will be to amplify the flux of an emission
line, with little distortion of the overall shape.  However, if the BLR
is a complex structure of transient knots within an overall turbulent
cloud region \citep{l06}, then provided the knots are of comparable size
or smaller than the Einstein disc of the lenses, individual hot spots of
random radial velocity will be microlensed to produce additional broad
emission lines with discrepant redshifts.

The effect of microlensing on the shape of quasar emission lines in
multiply lensed quasar systems is well illustrated by Keck
spectra\footnote {https://nexsci.caltech.edu/archives/koa}
of the wide separation quadruple lens SDSS J1004+4112 \citep{r04}.
Fig.~\ref{fig1} shows plots of the Si \textsc{iv} $\lambda1400$ and
C \textsc{iv} $\lambda1549$ broad emission lines for the A and B images
of SDSS J1004+4112, and in both emission lines the broad blue wing seen in
the spectrum of image A is replaced by a somewhat narrower red wing for
the B image.  Changes such as this are commonly found in spectrospic
surveys of gravitationally lensed quasars \citep{s12,g13b,f21}, and in
more detailed studies of individual systems \citep{e08,b14}.  As a result
of these studies, a few general points can be made about microlensing of
the BLR.  As mentioned above, if the entire emission region is
sufficiently compact, that is smaller than the Einstein disc of the
lenses, then microlensing will be seen as an increase in broad line flux,
but with little or no associated distortion of the line profile.  However,
for a larger and more complex structure of the BLR, individual knots with
non-systemic radial velocities can act as lenses to create composite
time-varying line profiles.  The change in shape of the Si \textsc{iv} and
C \textsc{iv} emission lines illustrated in Fig.~\ref{fig1} for
SDSS J1004+4112 would appear to be an example of this.

These observations raise the question of the timescale of structural
changes in the BLR.  In their analysis of broad line microlensing in
SDSS J1004+4112 \cite{r04}, as well as noting differences in line profile
between images A and B, also find that for image A the profile of the
C \textsc{iv} line changes by a large amount over a period of 6 months.
This very short timescale should be compared with the dynamical or cloud
crossing timescale \citep{p93} for the BLR in quasars, given by

\begin{equation}
\tau_{\rm{dyn}} = \frac{r}{V_{\rm{FWHM}}} \approx{50} \; \rm{years}
\label{eqn1}
\end{equation}

\noindent where $r$ is the radius of the BLR as measured in reverberation
mapping experiments, and is typically about a year for quasars
\citep{l18}, and $V_{\rm{FWHM}}$ is the Doppler width of the broad line
with a typical value of 6000 km sec$^{-1}$.  Given that by comparing the
spectra of different images it is established that the BLR is being
microlensed, it seems reasonable to conclude that the very short timescale
event in image A is also the result of microlensing.  This would be in
line with results from monitoring of the accretion disc of SDSS J1004+4112
for microlensing variations, where image C was observed to increase in
brightness by 0.7 magnitudes in 200 days \citep{h20b}.  The implication
here is that the accretion disc and BLR are being microlensed by the same
population of compact stellar mass bodies.

\section{Intrinsic variability of quasar broad emission line profiles}
\label{var}

\begin{figure}
\centering
\begin{picture} (150,250) (60,10)
\includegraphics[width=0.55\textwidth]{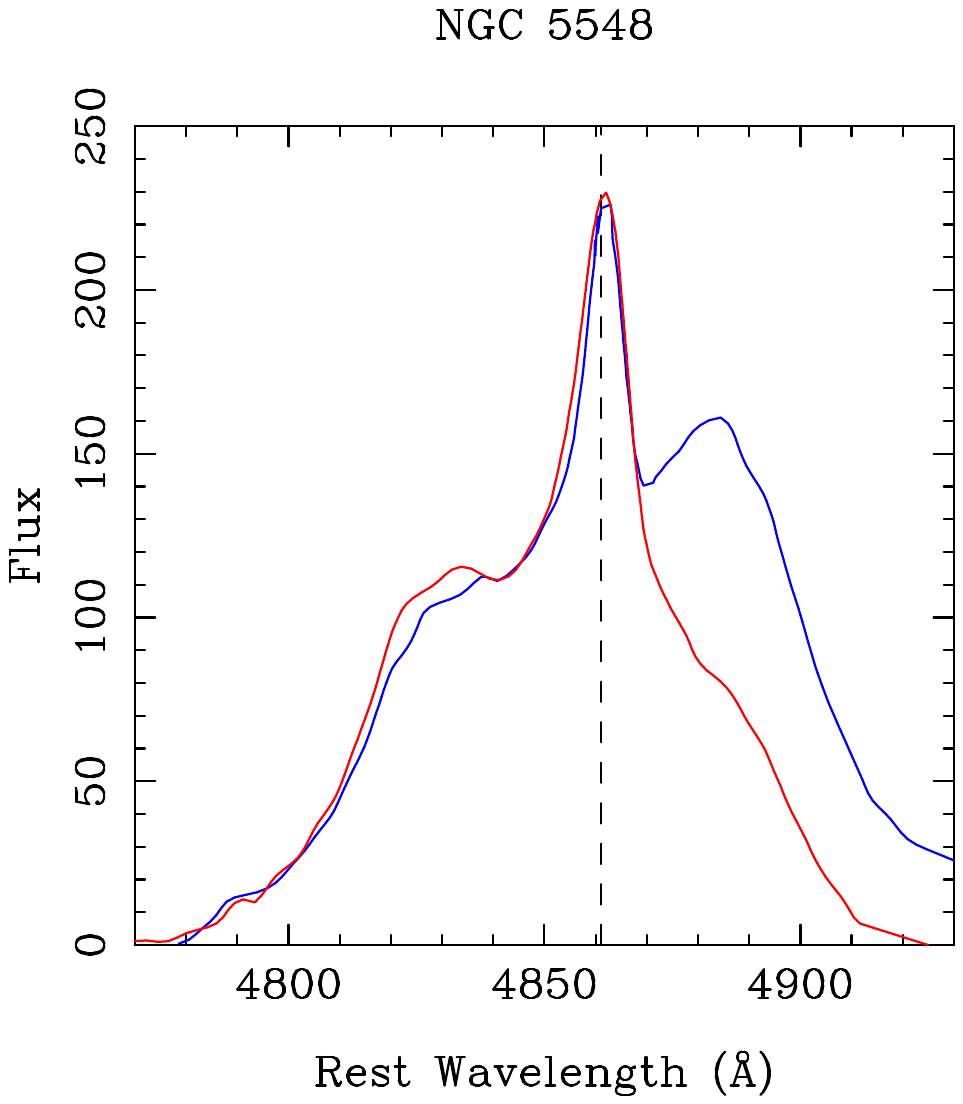}
\end{picture}
\caption{Emission line profiles for the H$\beta$ line in the Seyfert
 galaxy NGC 5548 from International AGN Watch data showing variations on
 a timescale of $\sim 5$ years.}
\label{fig2}
\end{figure}

It has for some time been well-established that changes in continuum
flux from the accretion disc of low luminosity Active Galactic Nuclei
(AGN) or Seyfert galaxies can produce variations in the strength of quasar
broad emission lines \citep{p02}.  In this case, the changes in emission
line strength typically follow the continuum flux changes by a few days,
representing the light travel time from the accretion disc to the broad
line clouds. This reverberation effect has also been observed by
\cite{k00} in relatively low luminosity quasars from the PG sample of
nearby quasars \citep{s83}.  More recently, reverberation mapping has been
extended to samples of luminous quasars \citep{k07,l18}, and some
interesting trends have emerged.  It is clear that as AGN become more
luminous, the associated BLR becomes more fragmented, and the response of
the emission lines to changes in continuum flux becomes more patchy.

Although most changes in emission lines can be explained by a simple
increase in brightness in response to changes in continuum flux from the
accretion disc, changes in the line profiles of Seyfert galaxies are also
observed on timescales of around 5 years \citep{w96}, consistent with the
dynamical timescale for Seyfert galaxies,
$\tau_{\rm{dyn}}  \approx{3-5} \; \rm{years}$ \citep{p01}.  These changes
in line profile are well illustrated in Figure 39 from \cite{p01}, and in 
Fig.~\ref{fig2} the superpositon of two H$\beta$ line profiles
from AGN Watch\footnote {http://www.astronomy.ohio-state.edu/~agnwatch}
data illustrates changes over the course of $\sim 5$ years.  However,
an important point to make is that there is no evidence for significant
changes in the shape or profile of the emission lines in response to
variations in the luminosity of the accretion disc.  This is consistent
with results from \cite{p99}, who conclude on the basis of an 8 year
survey that broad line profile changes are not reverberation effects, but
are due to mass motions within the BLR on the dynamical timescale.

\begin{figure}
\centering
\begin{picture} (150,250) (60,10)
\includegraphics[width=0.55\textwidth]{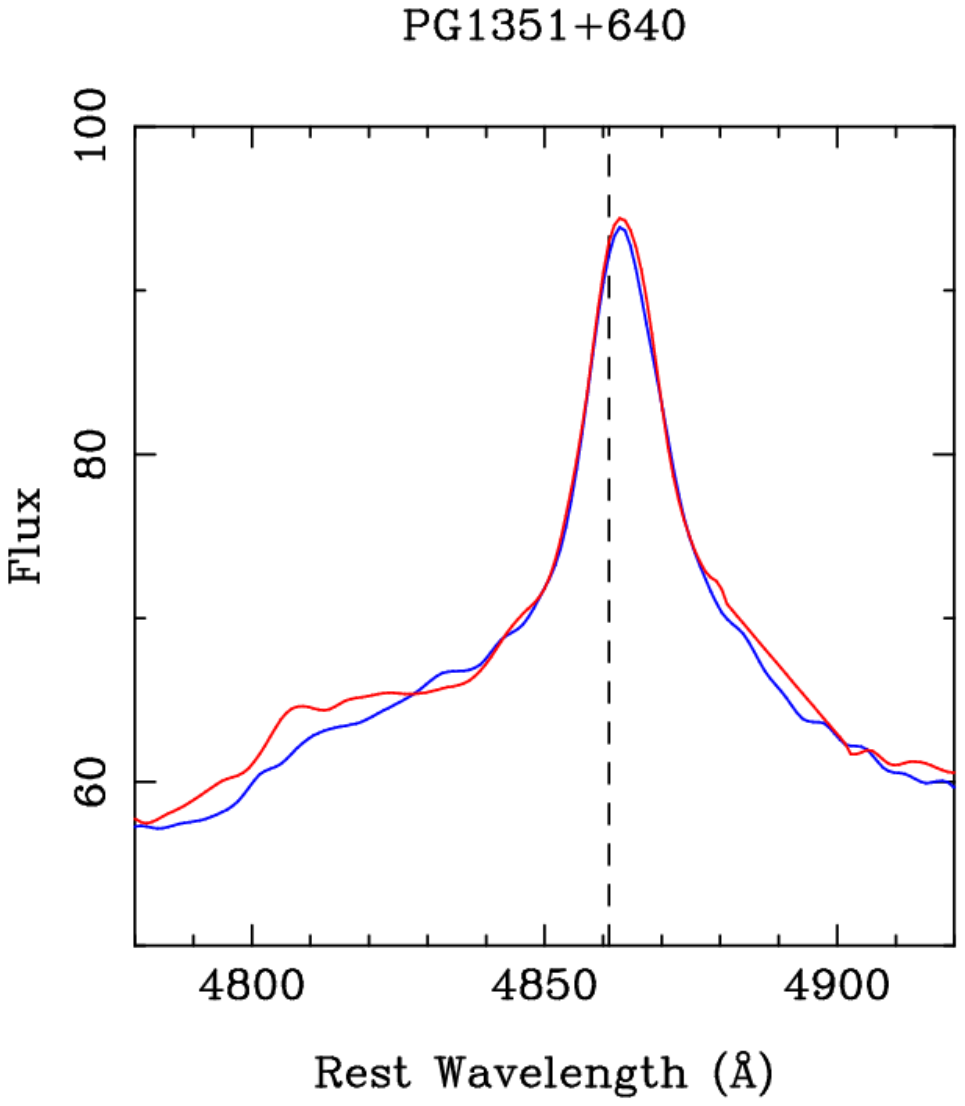}
\end{picture}
\caption{Emission line profiles for the H$\beta$ line of the low
 redshift quasar PG1351+640 from the spectroscopic monitoring programme
 undertaken by Kaspi et al. (2000).}
\label{fig3}
\end{figure}

Following on from the success of the International AGN Watch programme
\citep{p02} which focused on the Seyfert galaxy NGC 5548, attention turned
to investigating reverberation in more luminous AGN where the BLR is
expected to be larger.  This challenge was addressed by \cite{k00} with a
programme to monitor a well-defined sample of 28 Palomar-Green (PG)
nearby AGN \citep{s83} to look for reverberation effects in the broad
emission lines. The very large number of spectra which make up this survey
are available on line\footnote {http://wise-obs.tau.ac.il/$\sim$shai/PG/}.
Although it is true, as pointed out by \cite{k00}, that there were changes
in emission line profiles over the course of the 10 years of their survey,
examination of the spectra has shown that this only applies to the low
luminosity subset of their sample.  For quasars  with $M_B < -23$ and
reverberation timescales of around 200 days, the changes in emission line
shape are very small compared with those observed in Seyfert galaxies, as
shown in Fig.~\ref{fig2}.  This difference between changes in emission
line structure for high and low luminosity AGN is illustrated in
Fig.~\ref{fig3}, which shows the largest observed variation in emission
line shape for quasars over the 10 year monitoring period of PG quasars by
\cite{k00}.  It may be seen from Fig.~\ref{fig3} that the only difference
between the two profiles is a small enhancement of flux in the blue wing
of the H$\beta$ line, very different from the large changes in shape
illustrated in Fig.~\ref{fig2}.  This result is not surprising, as the
dynamical timescale for quasars or luminous AGN from Eq.~\ref{eqn1} is
of the order of 50 years, far longer than the 10 years of the PG
spectroscopic monitoring programme \citep{k00}.

\section{Microlensing in high redshift quasar spectra}
\label{hiz}

\begin{figure}
\centering
\begin{picture} (150,250) (60,10)
\includegraphics[width=0.55\textwidth]{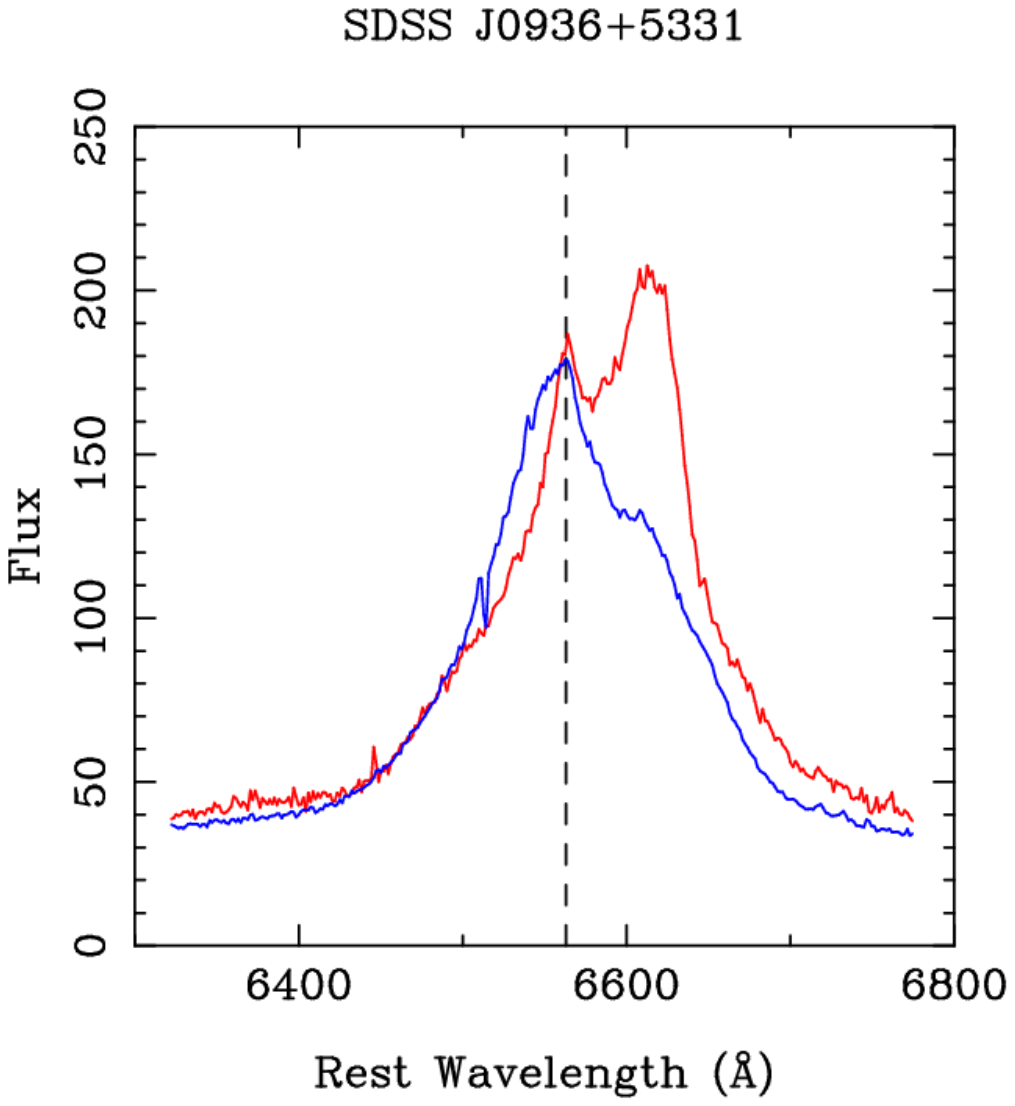}
\end{picture}
\caption{Emission line profiles for the H$\alpha$ line from SDSS archive
 spectra of the the quasar SDSS J0396+5331.}
\label{fig4}
\end{figure}

An interesting development in the study of emission line changes in quasar
spectra came from a search for evidence for binary supermassive black
holes in quasars \citep{l14}.  The investigation took the form of a
search for quasars with bulk velocity offsets in the broad Balmer lines
with respect to the systemic redshift of the host galaxy.  This resulted
in the compilation of a catalogue of 399 quasars from the Sloan Digital
Sky Survey (SDSS) with offset broad H$\beta$ lines, and a mean
redshift $z = 0.43$, around twice the mean redshift of the PG quasars.
Second epoch spectra of 50 of the candidates showed that for the most part
any changes in the spectra were limited to additional velocity offset,
with little change in the structure of the emission line profiles.
However, the authors flagged the case of SDSS J0936+5331 (illustrated in
Fig.~\ref{fig4}) which shows a strong additional red feature which
disappeared on a timescale of 10 years.  This short timescale suggests
that some other process unrelated to the dynamic timescale may be
involved, and provides a useful comparison with the PG quasars with a
mean redshift $z = 0.20$.

The possibility that the rapid changes in quasar emission line
profiles are associated with the presence of a binary supermassive black
hole (BBH) is addressed by \cite{l14}, who argue that most BBHs will not
exhibit double-peaked broad lines due to limitations in parameter space.
By implication, this rules out profile changes similar to those
illustrated in Fig.~\ref{fig4} for SDSS J0936+5331.  On this basis,
\cite{l14} rejected quasars with double peaked broad emission lines from
their sample of BBH candidates, which leaves open the question of what
physical processes are involved.

Further interest in changes in quasar emission lines was focussed
on the emergence or disappearance of broad emission lines in so-called
`changing look' (CLQ) quasars \citep{m16}.  Subsequent surveys for CLQs
were based on finding quasars for which large changes were observed in the
broad H$\beta$ line flux \citep{m19,g22}.  Between these two surveys some
30 new CLQs were discovered satisfying the adopted search criteria, which
meant that the recorded changes in the H$\beta$ line were confined to
total flux and not variations in line width or profile.  The first attempt
to find CLQs at high redshift $(z > 2)$ focussed on changes in the
C \textsc{iv} $\lambda1549$ emission line flux \citep{r20}, resulting in
the discovery of three quasars with large changes to the C \textsc{iv}
flux.  \cite{r20} also found the C \textsc{iv} profile to be approximately
constant, with the line flux responding to changes in continuum luminosity.

\begin{figure*}
\centering
\begin{picture} (0,200) (240,-10)
\includegraphics[width=0.32\textwidth]{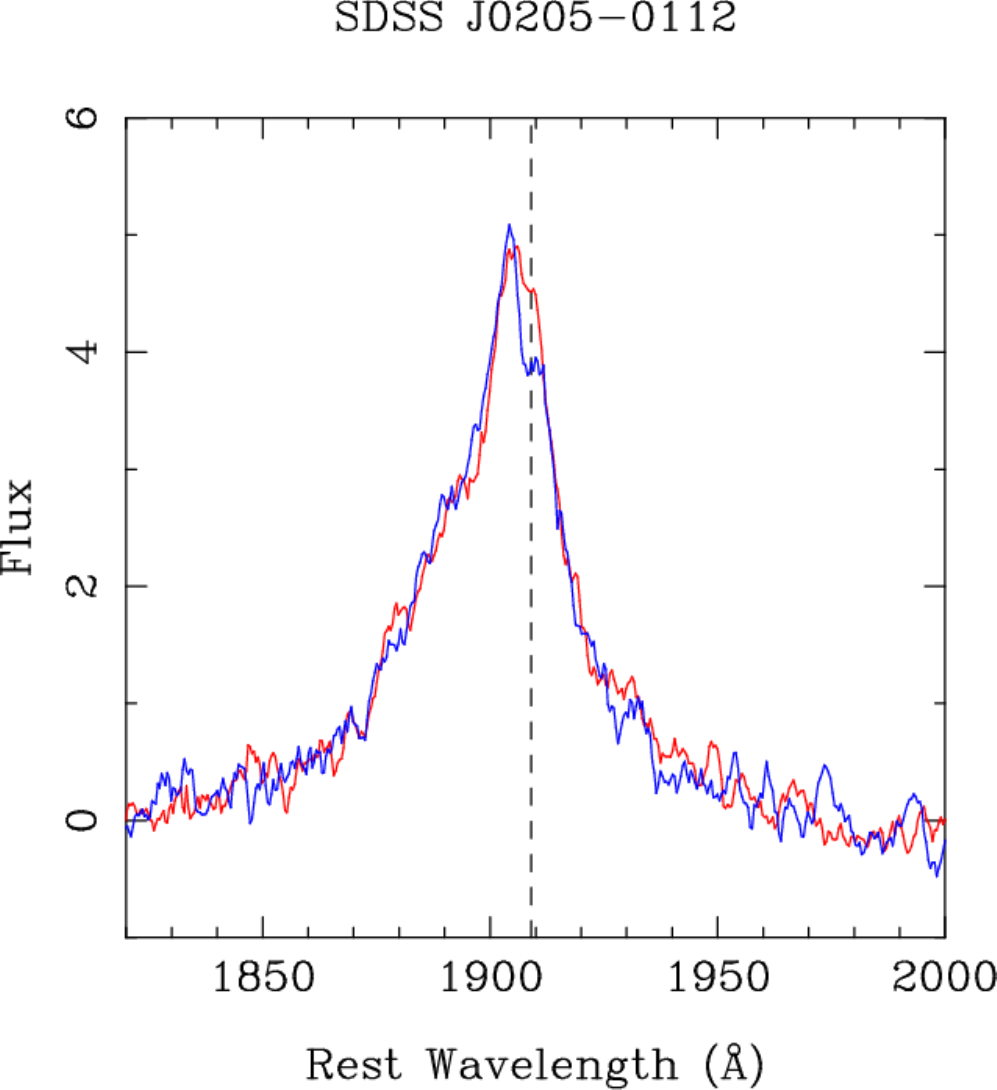}
\end{picture}
\begin{picture} (0,0) (70,-10)
\includegraphics[width=0.32\textwidth]{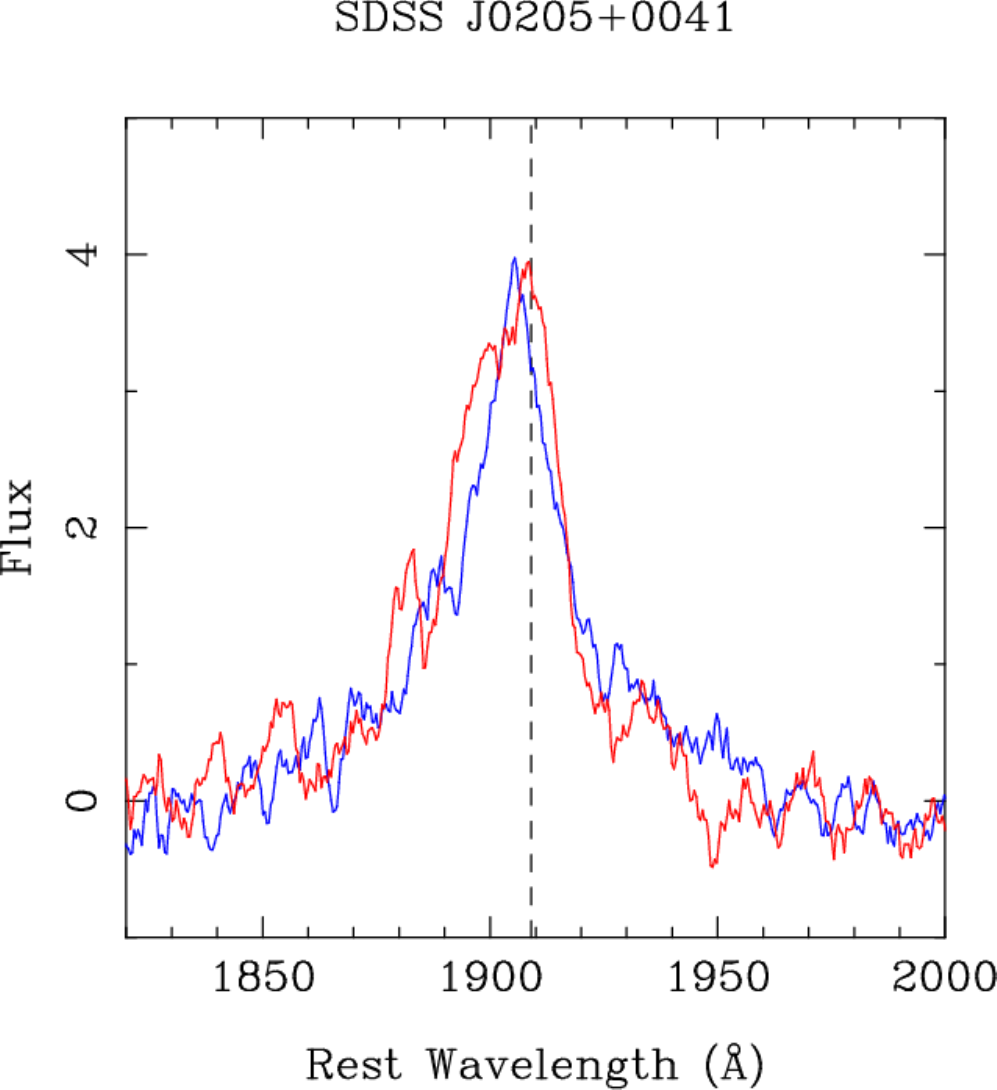}
\end{picture}
\begin{picture} (0,0) (-100,-10)
\includegraphics[width=0.32\textwidth]{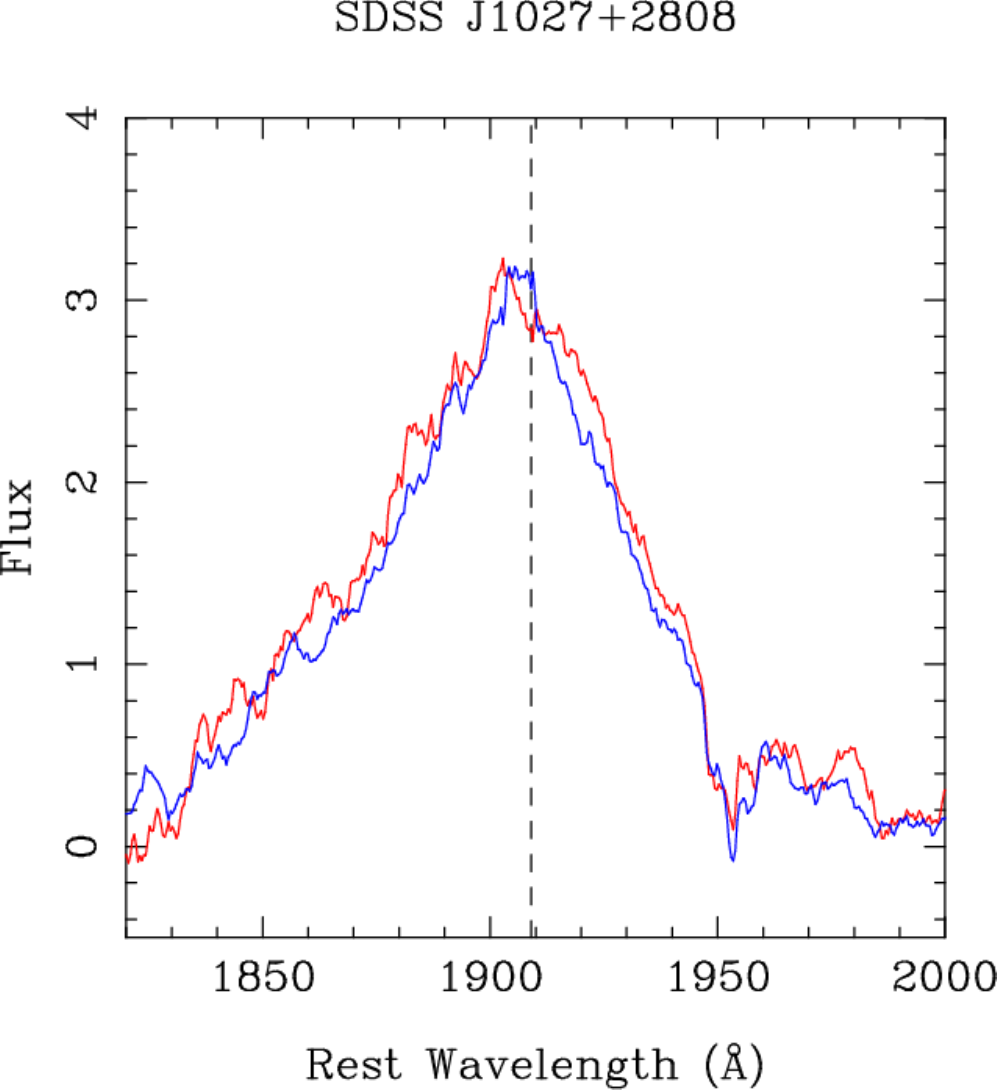}
\end{picture}
\caption{Emission line profiles for the C \textsc{iii}] line from SDSS
 archive spectra of quasars in the redshift range $1.5 < z < 3.0$, chosen
 for showing small changes in structure on a timescale of 10 years in the
 quasar rest frame.}
\label{fig5}
\end{figure*}

\begin{figure*}
\centering
\begin{picture} (0,600) (240,-410)
\includegraphics[width=0.32\textwidth]{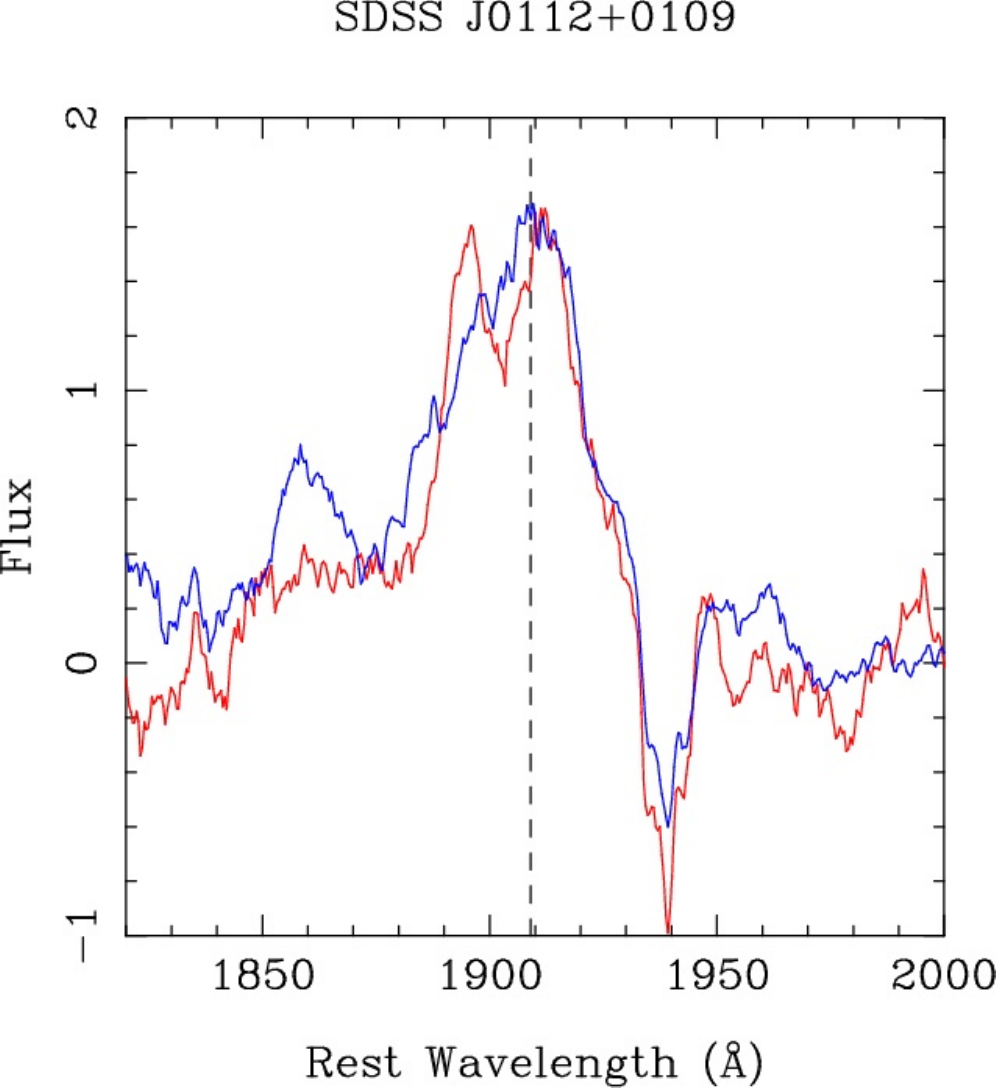}
\end{picture}
\begin{picture} (0,0) (70,-410)
\includegraphics[width=0.32\textwidth]{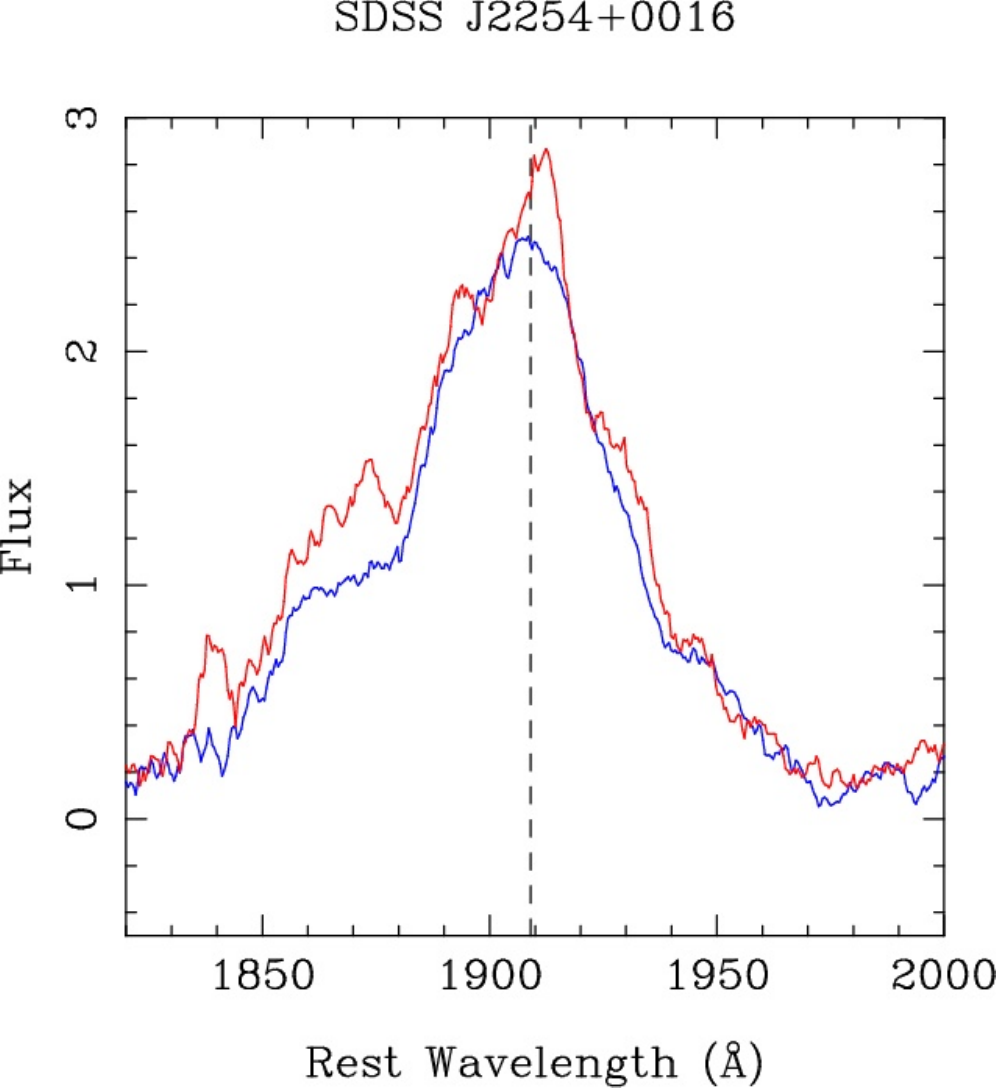}
\end{picture}
\begin{picture} (0,0) (-100,-410)
\includegraphics[width=0.32\textwidth]{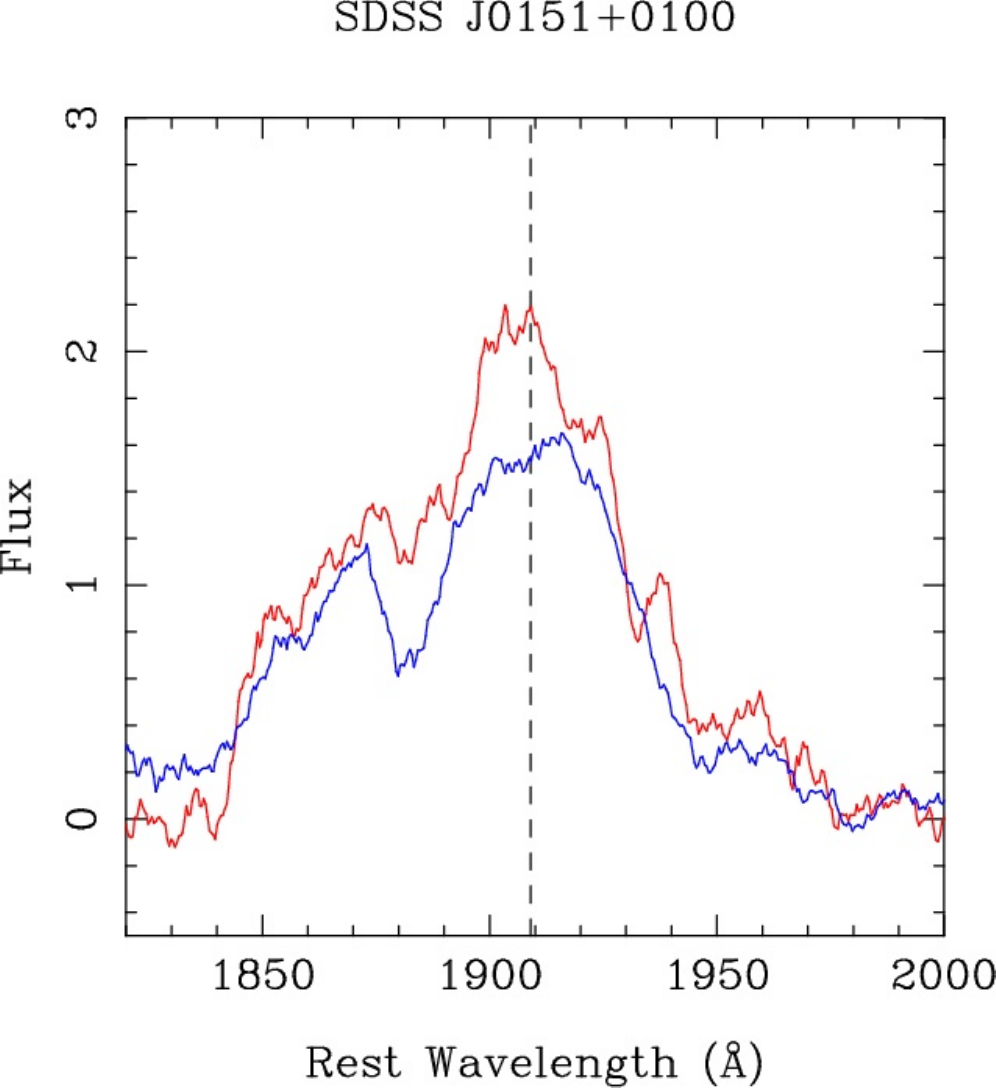}
\end{picture}
\begin{picture} (0,0) (245,-210)
\includegraphics[width=0.32\textwidth]{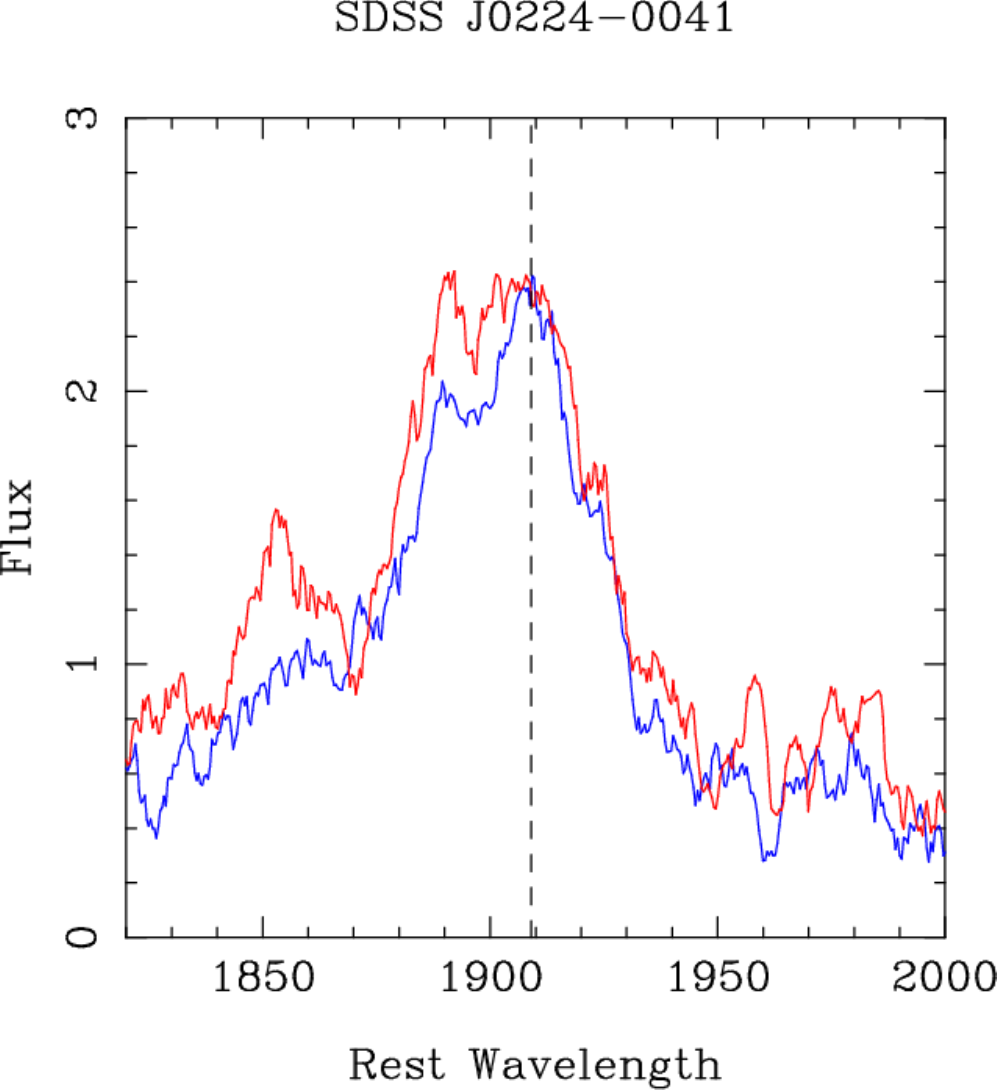}
\end{picture}
\begin{picture} (0,0) (75,-210)
\includegraphics[width=0.32\textwidth]{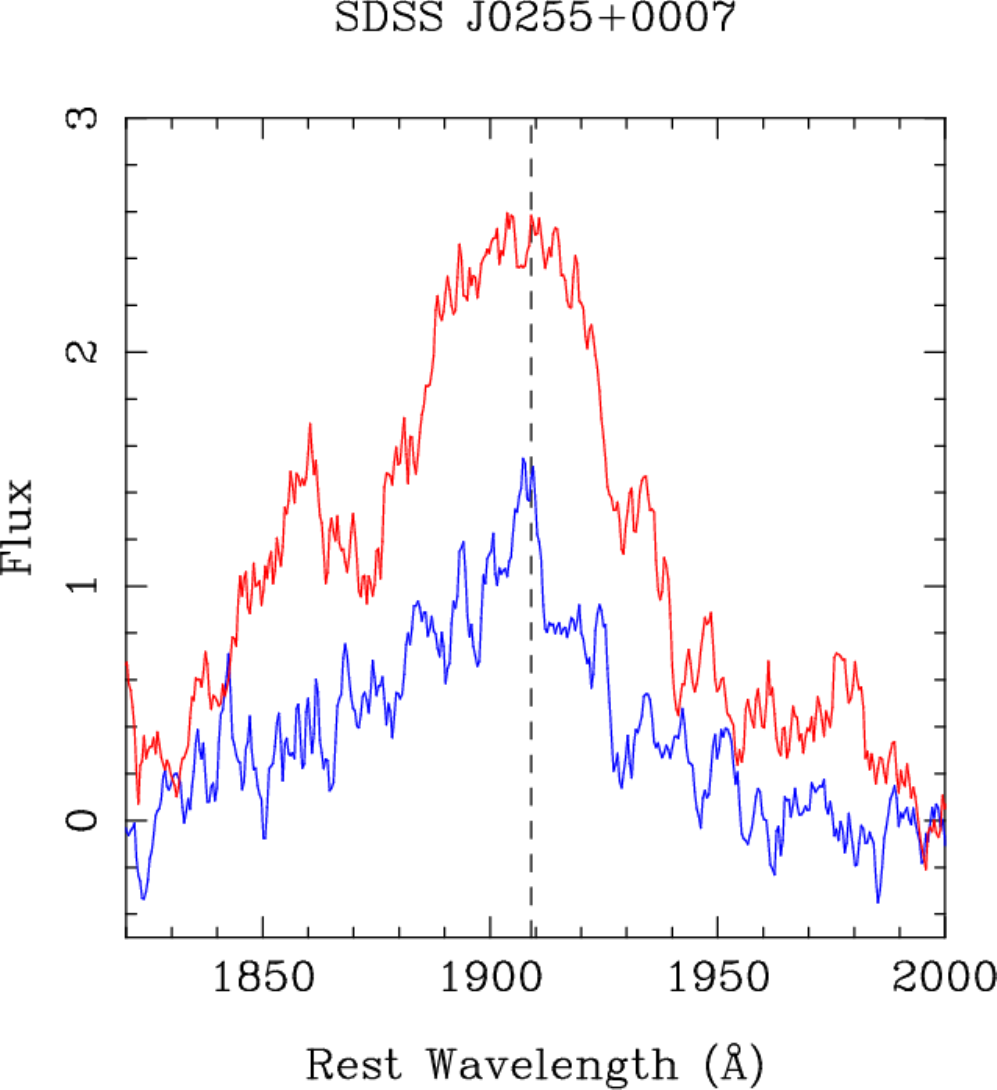}
\end{picture}
\begin{picture} (0,0) (-95,-210)
\includegraphics[width=0.32\textwidth]{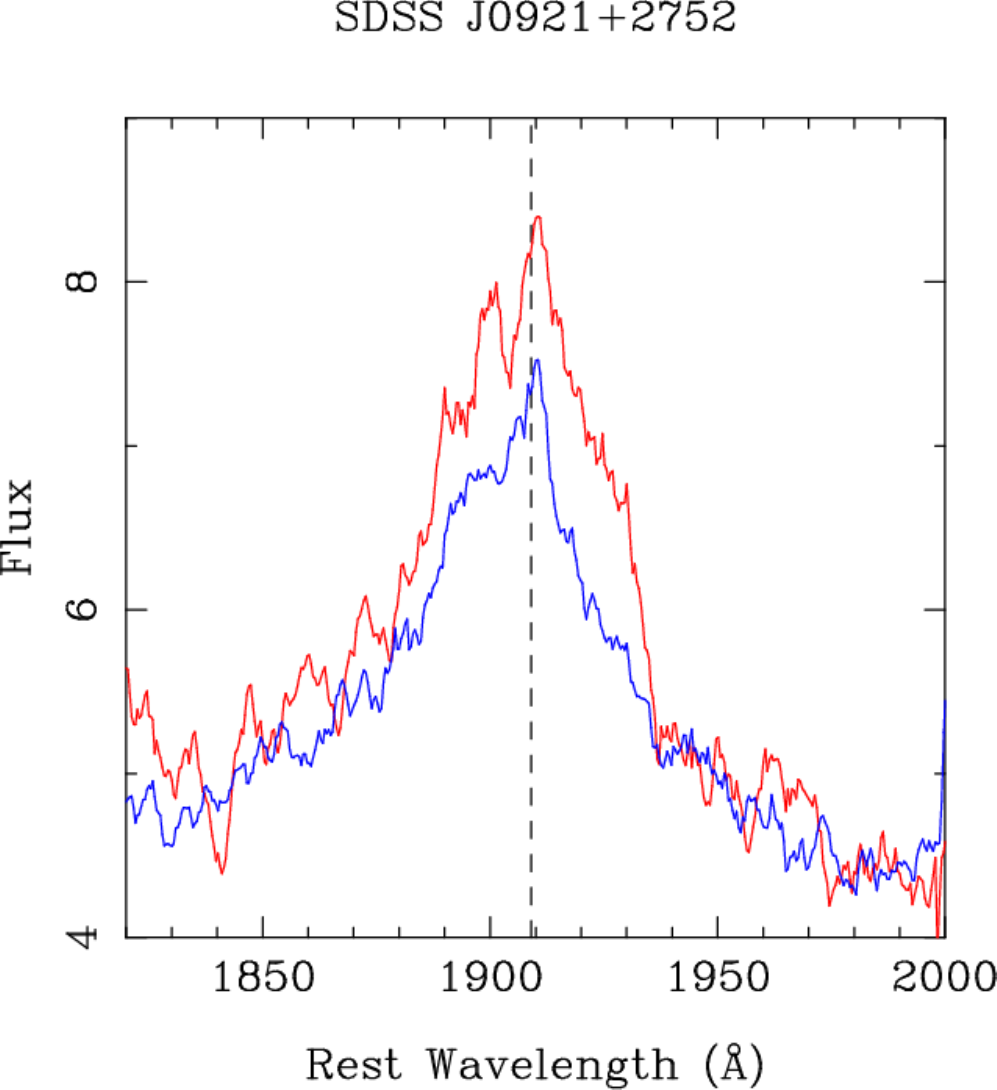}
\end{picture}
\begin{picture} (0,0) (250,-10)
\includegraphics[width=0.32\textwidth]{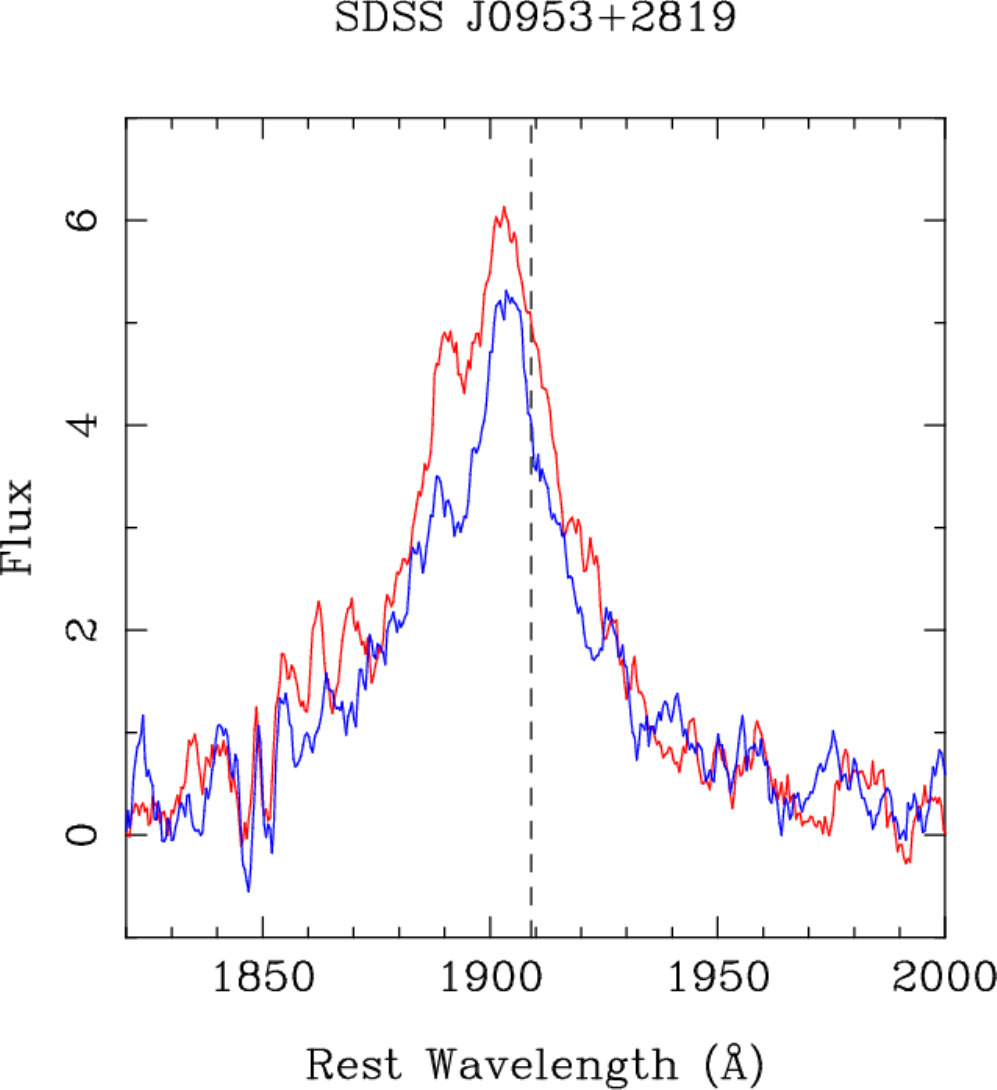}
\end{picture}
\begin{picture} (0,0) (80,-10)
\includegraphics[width=0.32\textwidth]{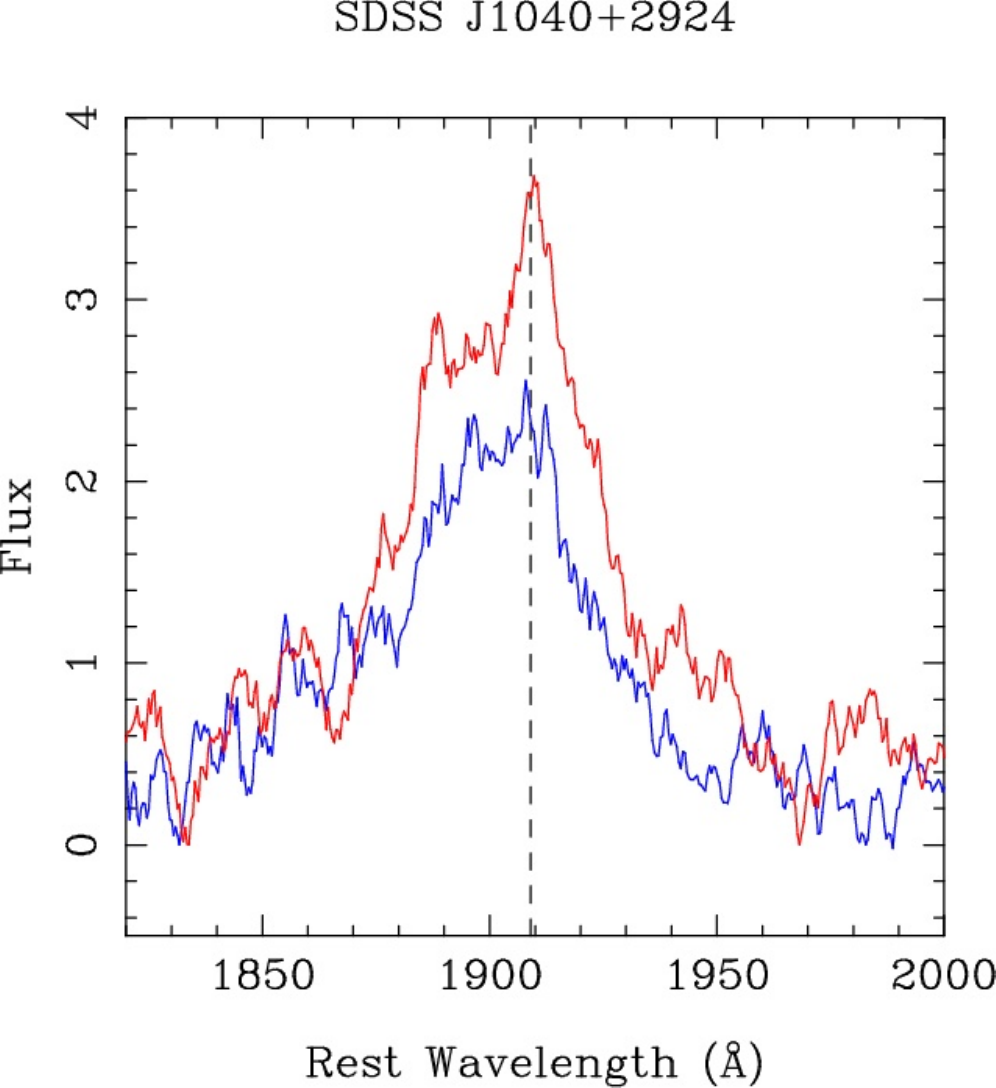}
\end{picture}
\begin{picture} (0,0) (-90,-10)
\includegraphics[width=0.32\textwidth]{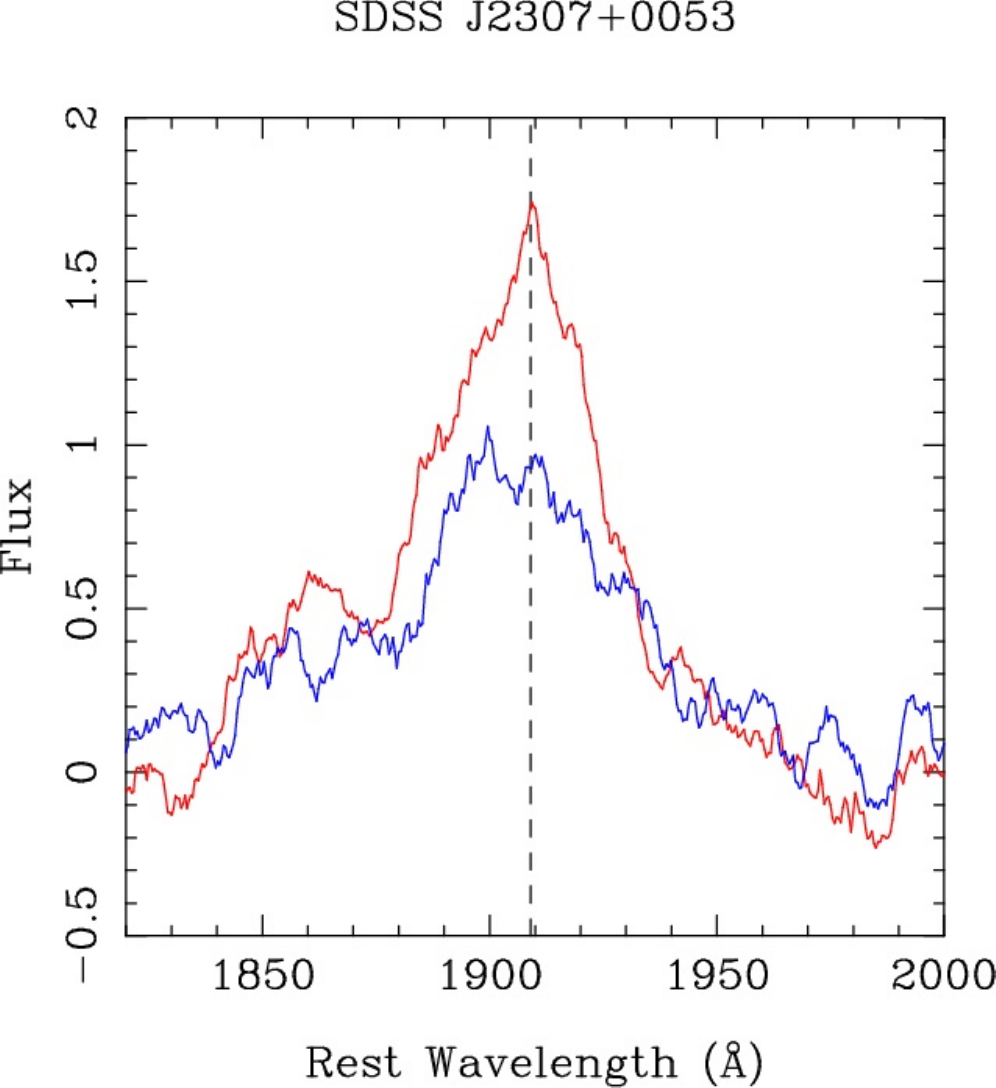}
\end{picture}
\caption{Emission line profiles for the C \textsc{iii}] line from SDSS
 archive spectra of quasars in the redshift range $1.5 < z < 3.0$, showing
 changes in structure on a timescale of 10 years in the quasar rest frame.}
\label{fig6}
\end{figure*}

The selection criteria adopted by \cite{r20} for their sample of high
redshift CLQs appear to leave open the question of the extent to which
changes in emission line profiles occur in the spectra of quasars with
$z > 2$.  To address this issue, the SDSS Time Domain Spectroscopic
Survey (TDSS)\footnote{
https://www.sdss4.org/dr17/spectro/extragalactic-observing-programs/tdss/}
provides a very good basis for selecting a sample of quasars for the
investigation of changes in emission line profiles. TDSS comprises a
number of sub-samples focusing on different categories of variable
objects, and of particular relevance for the study of quasar emission line
variability is the sample associated with the bitmask TDSS\_FES\_HYPQSO of
variable QSOs chosen for repeated observation.  In order to identify
profile changes in high redshift quasars, a subsample of the HYPQSO sample
was selected with redshift $1.5 < z < 3.0$, and the further requirement
that there were at least two spectra included in the HYPQSO sample for
each quasar.  This resulted in a final set of 53 quasars for further
study. Spectra of these candidates were then plotted out and examined for
obvious changes in emission line structure. The idea was not to compile a
complete sample, but to establish whether profile changes do occur in high
redshift quasars.  As expected, most of the quasar emission lines
showed little or no significant changes in emission profile.  As a general
rule the C \textsc{iii}] line lies in the part of the SDSS spectra with
the best signal-to-noise for the redshift range $1.5 < z < 3.0$, and
Fig.~\ref{fig5} illustrates three typical examples showing little or no
change of structure in the C \textsc{iii}] line.  However, detailed
examination of the spectra and superposition of emission lines from
different epochs revealed 9 quasar spectra, illustrated in
Fig.~\ref{fig6}, with unmistakeable changes in the C \textsc{iii}]
emission line profile.  These make up around 20\% of the sample and show
changes on a timescale of 10 years.  The changes typically take the form
of a broad emission line feature emerging in the blue wing of the
C \textsc{iii}] line.  Although it is possible that in some cases
such changes could be produced by misalignment of the object in the
aperture, where several observations are available new line structures
typically persist between two epochs separated by a short timescale.

The most natural explanation for the observed changes in line
structure is that they are intrinsic to the broad line region, resulting
from knots in a turbulent BLR emitting at non-systemic velocities in an
analagous way to that observed in Seyfert galaxies \citep{w96}, and
illustrated in Fig.~\ref{fig2} above.  Rapid changes in broad line shape
are only very rarely observed in low redshift quasars \citep{l14}, and
the dynamical timescale given in Eq.~\ref{eqn1} for luminous quasars would
seem to make such changes unlikely.  This appears to leave room for
a mechanism for variation external to the BLR for high redshift quasars
where rapid broad line changes are observed, and on the basis of the
discussion in Section~\ref{mic} above, microlensing of the BLR by a
population of stellar mass compact bodies would appear to be a
possibility.  It is also worth pointing out that the microlensing of other
blended lines may be responsible for the change in shape of
C \textsc{iii}].  In particular, Al \textsc{iii} lies close to the
appearance of new features in Fig.~\ref{fig6}, and microlensing of this
line may well contribute to the overall change in the structure of the
C \textsc{iii}] line.This of course does not mean that the broad emission
lines are being microlensed, but that if the conclusions of
\cite{h20a,h20b,h22} are correct, then the expected microlensing of quasar
broad emission lines is plausibly observed.

Rapid changes in emission line shape due to microlensing are well
illustrated by considering the differences in emission line profile due
to microlensing in the cluster lens SDSS J1004+4112, illustrated in
Fig.~\ref{fig1}.  The difference in light travel time between the two
images is very short, image B leading image A by 41 days \citep{f08},
which implies that the observed change in emission line structure cannot
be intrinsic to the emission line region.  After a careful study of
various possibilities \cite{r04} conclude that the observed variations in
emission line profile must be attributed to microlensing of part of the
broad line region of the quasar, resolving structure in the source plane
on a scale of $\sim \! 10^{16}$ cm.

As mentioned above, an important indication that high redshift quasars may
be microlensed is the contrast between line profile changes in low and
high redshift quasar samples.  Although the quasar sample of \cite{l14}
provides a useful low redshift sample for comparison, a more direct
control sample was obtained from the low redshift HYPQSO quasars.  In the
parent sample there were 10 members with redshift $z < 0.4$, corresponding
to a probability of microlensingg of around 1\%. Here the H$\beta$ line is
prominent in the high S/N part of the SDSS spectra, and the 10 quasars in
this subsample were examined in a similar way to the high redshift
objects, but in this case there was no evidence for significant changes in
emission line structure for any of the quasars.  This is consistent with
the results of \cite{l14} for changes in the H$\beta$ line for a sample of
50 low redshift quasars.

Despite the dynamical timescale given by \cite{p93} in Eq.~\ref{eqn1}
of around 50 years for intrinsic changes to broad emission line structure
in quasars, there are some caveats with the argument that this completely
excludes short term intrinsic variability of emission line structure.  For
example, a high ionization emission line blended with C \textsc{iii}] can
be globally magnified by intrinsic variability, implying that changes in
the broad emission lines of quasars can take place over relatively short
timescales and induce an apparent change in the shape of the
C \textsc{iii}] line \citep{s23}.  A related possibility is that regions
with different velocities can respond with different time lags to
intrinsic variability causing differences in the shape of emission line
profiles observed at different epochs \citep{g13a,d18}.

\cite{r04} appeared to accept the widely held view at the time that the
compact bodies acting as microlenses were stars in the lensing cluster.
Given the apparent absence of starlight at such a great distance (60 kpc)
from the cluster centre of SDSS J1004+4112, this somewhat implausible
hypothesis was examined in detail by \cite{h20b}. The result of this
study, as dicussed above, showed that from measurements of starlight in
the vicinity of the quasar images, the optical depth to microlensing from
stars was far too small to explain the observed differential changes in
brightness of the quasar images.  The main conclusion was that to account
for the observed microlensing, the lenses must make up at least a large
part of the dark matter.  This conclusion was reinforced by a more general
study of quasar variability on a cosmological scale \citep{h22}, where it
was concuded that to account for the distribution of quasar lightcurve
amplitudes it was necessary to include the microlensing effects of a
cosmologically distributed population of stellar mass compact bodies.
These results suggest that such a population of compact bodies might also
be responsible for microlensing the broad line regions of quasars in the
general field, and thus produce the observed rapid changes in broad line
profile.

\section{Discussion}

\begin{figure}
\centering
\begin{picture} (0,150) (120,165)
\includegraphics[width=0.48\textwidth]{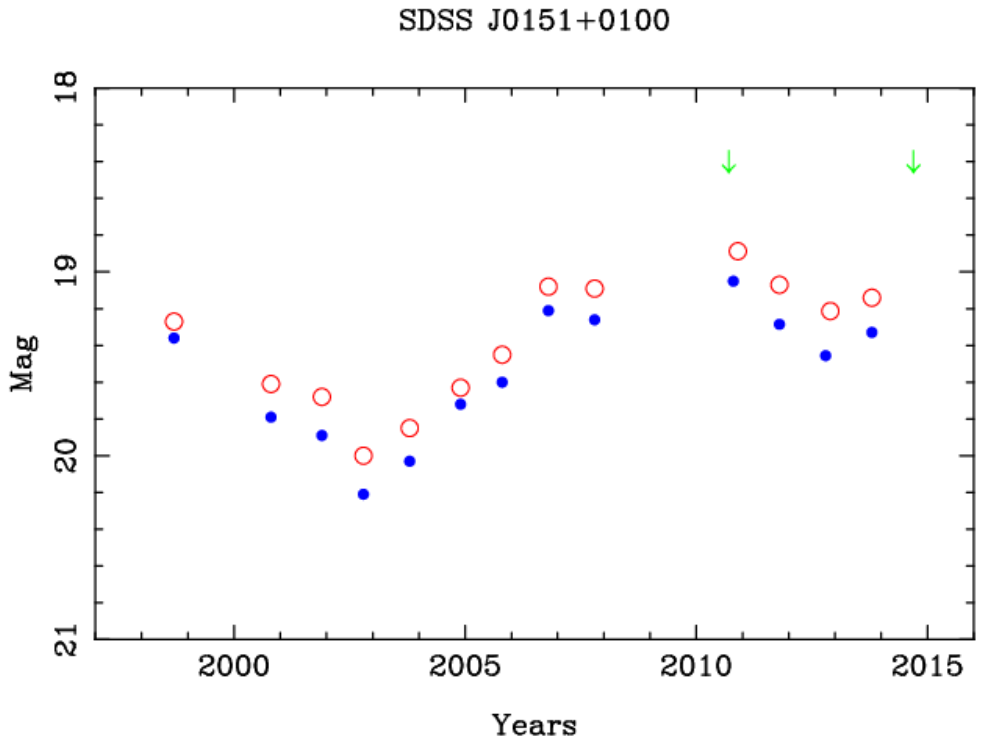}
\end{picture}
\caption{Light curve for the quasar SDSS J0151+0100 in the g-band (blue
 filled circles) and the r-band (red open circles).  Data for the years
 1998 to 2007 are from the SDSS stripe 82 archive, and for 2010 to 2013
 from the Pan-STARRS1 data archive.  The green arrows mark the epochs of
 the two spectra shown in Fig. 7}
\label{fig7}
\end{figure}

The idea behind this paper has been to review evidence that quasar broad
emission lines are being microlensed.  For this to occur, there must be
a substantial optical depth to microlensing $\tau$ of lenses along the
line of sight to the quasar.  For quasars in multiply lensed systems there
remains the possibility that the population of lenses is associated with
the dark matter in the lensing galaxy or cluster halo, but for isolated
quasars in the general field where the lenses are assumed to make up the
dark matter, the expectation of microlensing events will depend on the
redshift of the quasar.

It has been demonstrated above that low luminosity AGN or Seyfert galaxies
at low redshift show changes in emission line profile on a timescale of a
few years.  This is compatible with the dynamical timescale of 3-5 years
for such small broad line regions, and there is no need to invoke
microlensing which would be very unlikely at such low redshift with a
corresponding low value of $\tau$.  On the other hand, the low redshift
sample of quasars show no significant changes in emission line profiles.
These quasars are certainly too nearby for there to be any significant
chance of microlensing, and the dynamical timescale for the large
associated BLR means that any intrinsic changes to the emission line
profiles would occur on a timescale of $\sim \! 50$ years, far longer than
the length of the monitoring programme \citep{k00}.  There are
however some caveats to this broad picture.  A high ionization line such
as Al \textsc{iii} blended with the C \textsc{iii}] line can reverberate
differently to changes in continuum flux, which can occur on short
timescales.  The resulting changes in the flux ratio from the two lines
centered on different wavelenghts can thus plausibly result in significant
changes in broad line structure over a relatively short timescale.
Another possibility is that regions with different velocities can respond
with different time lags to changes in the continuum source.  This can
then result in apparent changes in the structure of the broad line region
when observed at different epochs.

A new approach to changes in quasar broad lines was proposed by
\cite{l14}, with the focus on measuring bulk offsets in line velocity
relative to the systemic redshift of the quasar.  The idea was to look for
evidence of binary supermassive black holes in quasars, but their data
have turned out to be very useful for studying changes in quasar emission
line profiles.  In the sample of 50 candidates selected by \cite{l14},
only the quasar SDSS J0936+5331 showed unmistakeable evidence for changes
in the H$\beta$ line profile, with a marginal additional candidate
SDSS J1345+1144.  Given that the target emission line was H$\beta$, the
typical redshift of the sample members was inevitably small, with a median
value $z \approx 0.4$ corresponding to an optical depth to microlensing
$\tau = 0.01$ in a standard $\Lambda$CDM Universe \citep{f92}.  This
implies a probability of microlensing for an average sample member of
$\sim 1\%$, which is not inconsistent with the microlensing of just one
sample member. However, the important thing to note is that there is no
evidence for widespread changes in emission line profile in such a low
redshift sample. This is consistent with the long dynamical timescale
expected in quasars for changes in the structure of the broad line region,
as well as the small probability of microlensing in such a low redshift
sample.

The question of emission line profile changes in high redshift
\mbox{$(z > 2)$} quasars was first addressed by \cite{r20}.  Their
candidates were selected from the SDSS archives on the basis of optical
variability and the availability of a second spectrum.  Subsequent visual
inspection revealed 3 quasars showing interesting emission line behaviour,
and follow-up spectroscopy confirmed the changing nature of the emission
lines, with particular focus on the Ly~$\alpha$, C~\textsc{iv} and
Mg~\textsc{ii} lines.  The main changes observed in the emission lines
were line emergence and collapse, but no evidence was reported for
significant changes in the shapes of the line profiles.  The authors
further conclude that the main driver for emission line variability is
the broad-band continuum itself, but declare themselves `agnostic' as to
the underlying physical processes.

An outstanding question remaining after the investigation by \cite{r20}
is whether structural changes in broad emission line profiles can occur in
high redshift quasar spectra on timescales of a few years, as occasionally
observed in low redshift quasar samples \citep{l14}.  To answer this
question, the HYPQSO sample of quasars earmarked for repeated observations
has provided a useful start.  Given the rarity of changes in emission line
profiles for low redshift quasars \citep{l14}, where emission line changes
are largely confined to increase or decrease in line flux with no
associated change in the shape of the line profile, it was surprising to
find that profile changes in the HYPQSO sample with $1.5 < z < 3.0$ were
readily identified, as illustrated in Fig.~\ref{fig6}.  The implied
dependence of changes in line profile on redshift rather than BLR size
suggests that some external mechanism may be involved.  Given that compact
bodies are well known to microlens broad emission lines in multiply imaged
quasar systems \citep{r04,s12,f21}, microlensing of the broad emission
lines may provide a solution to the observed changes in emission line
structure in luminous quasars.

The idea that quasar emission lines are being microlensed raises some
immediate questions.  Firstly, what is the origin of the lenses?  This
can be answered on the basis of the original motivation for this
paper, which arose from evidence \citep{h20a,h20b,h22} that microlensing
observations implied at least a major component of dark matter to be in
the form of stellar mass compact bodies, most plausibly primordial black
holes.  One of the consequences of this would be the microlensing of the
BLR in quasars of sufficiently high redshift, thus providing an
explanation for the observed changes in emission line profiles.  To put
this on a firmer statistical basis the probability of microlensing can be
estimated from the optical depth to microlensing $\tau$.  Using the
equations of \cite{f92}, the value of $\tau$ for a quasar with $z = 2$ in
a $\Lambda$CDM Universe is $\tau \approx 0.2$, implying a probability of
around 20\% that the quasar will be significantly microlensed.  This is
interestingly close to, and certainly consistent with, the proportion of
quasars found to show changes in emission line structure for the TDSS
subsample described in Section~\ref{hiz}.  In fact the probability is
somewhat higher than this, as for values of $\tau \gtrsim 0.1$, the
amplification patterns of the individual lenses will start to combine in
a non-linear way to form a pattern of high-amplification caustics
\citep{k97}.  However, at a redshift $z \sim 2$ this effect will be small.

By a fortunate circumstance, one of the high redshift quasars described
in Section~\ref{hiz} and illustrated in Fig.~\ref{fig6}, where changes in
emission line profile were detected, was also included in the SDSS Legacy
photometric monitoring programme in Stripe 82\footnote
{https://faculty.washington.edu/ivezic/macleod/qso-dr7/Southern.html},
providing the opportunity to look for any unusual features in the light
curve which might be connected with changes in emission line profile.
The SDSS lightcurve for SDSS J0151+0100 is plotted in Fig.~\ref{fig7},
together with additional measures from the Pan-STARRS1 data archive
\footnote{https://catalogs.mast.stsci.edu/panstarrs/}\citep{f20}.  The
light curve shows achromatic variation with amplitude of a magnitude over
a timescale of around 10 years.  There is no indication of short term
fluctuations in brightness which might be associated with changes in
emission line strength.
 
\section{Conclusions}

This paper has set out to investigate whether changes in the structure of
quasar broad emission lines are consistent with the expected microlensing
by a population of stellar mass primordial black holes making up a large
fraction of dark matter.  Evidence for such a population has been
published recently \citep{h20a,h20b,h22}, which raises the possibility
that these compact bodies may also be detected by microlensing the broad
emission line clouds associated with quasars.  Microlensing effects are
expected to change the shape of the broad line profile, typically
resulting in the appearance of new emission features offset from the
systemic velocity of the quasar.  Although such changes can happen as a
result of random motions of the clouds in the broad emission line region,
they would be expected to occur over very long dynamical timescales of the
order of 50 years or more, as opposed to microlensing timescales of around
5 years.  A further discriminant comes from the redshift of the quasar.
Low redshift quasars are very unlikely to be microlensed due to the
predicted small optical depth to microlensing of the lenses, but for
redshift $z > 2$ the probability of microlensing rises to around 20\%.

The main results of the paper are as follows:

\begin{enumerate}
\item The paper starts by characterising the changes in quasar emission
line structure due to microlensing, based on spectroscopy from the
literature of quasar spectra from gravitational lens systems where
microlensing is known to occur.
\item Changes in emission line structure in low redshift Seyfert galaxies
are illustrated, and attributed to cloud motions in the BLR.  These
changes occur on short timescales of the order of 5 years, corresponding
to the dynamical timescale of the BLR.
\item Luminous low redshift quasars show no changes in emission line
structure.  This is attributed to the much larger size of the BLR, with a
dynamical timescale of the order of 50 years.
\item Luminous quasars at high redshift $(z > 2)$ exhibit different
behaviour, with around 20\% showing strong changes in emission line
structure on a timescale of 5 years.  This is consistent with the expected
microlensing from a population of solar mass compact objects comprising at
least a large fraction of the dark matter, although intrinsic
variability is an alternative explanation.  The most plausible candidates
for the compact bodies are primordial black holes.

\end{enumerate}

\section*{Acknowledgements}
This research has made use of the Keck Observatory Archive (KOA), which
is operated by the W.M.Keck Observatory and the NASA Exoplanet Science
Institute (NExScI), under contract with the National Aeronautics  and
Space Administration.

\section*{Data Availability}

The data upon which this paper is based are all publicly available and are
referenced in the text, with footnotes to indicate online archives where
appropriate.


\bsp	
\label{lastpage}
\end{document}